\documentstyle[12pt]{article}

\begin{document}
\date{September 1993}
\title{Modular Theory and Symmetry in QFT}
\author{B. Schroer\\ Institut f\"ur Theoretische Physik\\
Freie Universit\"at Berlin}
\maketitle
\begin{abstract}

The application of the Tomita-Takesaki modular theory to the Haag-Kastler net
approach in QFT yields external (space-time) symmetries as well as internal
ones
(internal ``gauge para-groups") and their dual counterparts (the ``super
selection
para-group"). An attempt is made to develop a (speculative) picture on
``quantum symmetry"
which links space-time symmetries in an inexorable way with internal
symmetries.
In the course of this attempt, we present several theorems and in particular
derive the
Kac-Wakimoto formula which links Jones inclusion indices with the asymptotics
of expectation values in
physical temperature states. This formula is a special case of a new asymptotic
Gibbs-state
representation of mapping class group matrices (in a Haag-Kastler net indexed
by
intervals on the circle!) as well as braid group matrices.
\end{abstract}
\newpage

\section{Introduction}

The topics to be discussed in this paper belongs to the so-called algebraic
QFT.
This terminology is referring to the mathematical methods used in the approach.

But the methods are not the most characteristic feature of this theory. It
could
incorporate any mathematical methods consistent with relativistic quantum
physics.

A more intrinsic characterization would be to say that it is founded on
mathematically
formulated physical principles [1].

There is another better known approach based on functional integrals,
which resulted from a formal extension of perturbativly successful quantization
ideas
(founded on the presumed existence of a classical-quantum parallelism i.e. a
kind
of inverse of the Bohr correspondence principle).

Algebraic QFT is in some physical sense very closely
linked with the older ``dispersion theoretical approach" whose modest success
culminated
in the theoretical derivation of Kramers-Kronig dispersion relation for the
scattering amplitudes of relativistic particles
from the two
main principles of relativistic local quantum physics [1]: Einstein causality
and spectral
properties (Dirac stability by filling negative energy states.) as well as
their subsequent
experimental verification in the case of p-p forward scattering.

Algebraic QFT
uses the same physical principles, but employs a vastly more subtle
mathematical
framework (von Neumann Algebras, Jones Inclusion Theory etc.) than just
Cauchy's theorem in dispersion
relation.

A rough look at those old principles reveals that they
can be divided into two groups as mentioned above: causality properties
expressed in terms of
commutativity properties of local algebras and stability properties expressed
in terms of states on observable algebras (or to be a little bit more precise,
in terms of spectral or modular (KMS) properties of representation of algebras
canonically (GNS) affiliated with these states). There are also related
properties
which join algebras and states as e.g. the notion of ``statistical
independence" of
quantum physics in space-like seperated regions leading, among other things, to
limitation
on multiplicities of high energy states and yielding the existence of
temperature
states from the knowledge of the vacuum states [2].

Since in no other area of theoretical physics (e.g. string theory, 2-d
quantum gravity) anybody explains in detail the principles on which his
approach
is founded, I will follow this general trend. In my case, these are however not
representing an
``industrial secret"  but they have been exposed nicely in the literature [3].

The main idea
is to avoid making assumptions on charge-carrying fields but rather to define
them and derive their properties from observables. (In that respect algebraic
QFT
is even different from QFT in the spirit of Wightman).

Algebraic QFT has been very successfull
at that. It is e.g. quite easy to define a composition of charged
localized states on the
observable algebra $\cal A$ (which is really a net of algebras):
\begin{equation}
\omega_1, \omega_2 \longrightarrow \omega_1\times\omega_2 (A) \not=
\omega_2\times\omega_1(A) \qquad A\epsilon{\cal A}
\end{equation}
which commutes (the equality sign holds) if the localization is space-like
i.e.\hfill\break
 $loc\ \omega_1$ {\LARGE\sf x} $loc\ \omega_2$.
The statistics operator $\varepsilon$ shows up if one tries to lift states to
state
vectors. For spacelike separations on encounters the so called statistics
operator $\varepsilon$
as an obstruction [4].
\begin{equation}
\psi_{\omega_2\times\omega_1} = \varepsilon \psi_{\omega_1\times\omega_2}
\end{equation}
 One immediatly realizes the analogy to the Wigner phenomenon of ray
representations of
symmetry groups which arise if one lifts the algebraic automorphisms to  the
space of vector states or of Berry's phase (the latter being of a
more phenomenological nature) which is a well-known measurable obstruction of
atomic physics. Statistics is (with the exeption of the abelian case
the so called anyons) not described by a phase, but a unitary operator.

With a bit
of additional work one finds that the operator $\varepsilon$ generates the
braid group
(special case is the permutation group) in low dimensional QFT [5] and the
permutation group
otherwise [6]. Furthermore there are Markov-traces leading (after rescaling)
to knot invariants, and by considering
 ``$\infty$" knots (i.e. suitable limits resembling infinite order perturbation
theory
for vacuum polarization in QFT), one obtains invariants of 3-manifolds [7].

There are however two physically important areas which have remained rather
difficult
territory even for algebraic QFT. Unfortunatly these are just the areas in
which the
new low-dimensional theoretical discoveries could be confronted with
experimental
reality (e.g. of condensed matter physics).

One is the problem of constructing and classifying
the simplest fields which obey the new statistics in the ``freest" way as it is
allowed
by braid group statistics i.e. the analogon of free bosons and fermions (where
the statistics refers to the permutation group) which may be called ``free
anyons" (abelian)
and more generally ``free plektons" (nonabelian).

In conformal field theory, according to our present best
understanding, only free objects (but they do not describe particles but rather
``infraparticles")
exist.

The second important physical problem is symmetry. In the old theories of
fermions and
bosons one had a sharp distinction between external (space-time) and internal
(isospin, flavour) symmetries. Any attempt to ``marry" them in a nontrivial
way was doomed
by failure [8].

In the new theory to the contrary there are obstructions against a
complete divorce e.g. in conformal field theory some global conformal
transformations
($2\pi n$ rotations on $\tilde S^1$) are inexorably  linked with internal
charge transformations [7].

According to our present best understanding of 3-d theories with braid group
statistics, it is perfectly conceivable that those already mentioned limiting
invariants of 3-manifolds
show up in some new perturbation theories of plektons in infinite order [7].
With
other words curved 3-manifolds could appear in ordinary (infinite order)
plekton scattering (i.e. even though
the ``living space" of the 3 dim. theory remains Minkowskian). In this case the
3-manifolds
are expected to have something to do with a  ``space" formed by external
charge-type
variables together with (subgroups of ) the M\"obius-group, resp. the 3-d
Poincar\'egroup: Transl. $\bowtie 0(2,1)$.

The so called Topological Field Theory [10] in itself offers no concept of a
physical ``living space" of fields i.e. the localization necessary for a
physical interpretation is
absent and therefore the physical interpretation of its topological content
is difficult and ambiguous which, however in no way limits its mathematical
usefulness.

These problems with the physical interpretation can be traced back to the
global
nature of the Feynman-Kac functional integral representation which (from the
point of
view of general QFT) constitutes an attempt to simultaneously guess a state and
an
operator algebra (e.g. its vacuum-representation) on the basis of perturbative
experience. Causality- and locality-properties can only be checked at the end
after a complete GNS reconstruction to the real-time analytically continued
expectation values has been carried out.
The formal locality properties of the classical action are  necessary (for
quantum locality),
but by no means sufficient.
This is exemplified by the use of non-regular states on nets of Weyl algebras,
\footnote{The formal canonical quantization of Chern-Simons actions yields
Weyl-like algebras on 1-forms as their operator-algebraic content.}
 which convert the
``would be" translationally covariant $III_1$ factor into $II_1$
hyperfinite tracial factors as used by V. Jones
(and the model becomes in this way ``topological"):
 all continuous space-time covariances
become ``knocked out" by the singular (local gauge invariant) states [11], i.e.
are
not unitarily implementable in a continuous fashion.\footnote{This singular
state
description is presently only known for the abelian Chern-Simons action
[11]. For illustrating the above warning, this is sufficient. In quantum
mechanics
such singular states destroy the continuous translation and violate the
uniqueness
theorem of the Schr\"odinger representation.}

A study of singular states on suitably defined nonabelian Weyl algebras over
1-forms in 3d which could explain the tantalizing numerical coincidence [62] of
the
semiclassical approximations in the continuous Witten approach with the exact
Turaev Viro combinatorical theory (which is identical to that of algebraic QFT
if
one specializes to models with 2-channel (Hecke-Jones) and 3-channel
(Bierman-Wenzl) statistics)
 and yield also the correct pair $\{$algebra, state$\}$
behind Topological Field Theory is still missing. However I am not aware of any
alternative idea which captures the intuitive picture of averaging over ``gauge
copies" in operator language (i.e. without resorting to semiclassical or
quantization ideas as the BRST formalism).

Singular states have first been applied in QFT by Buchholz-Fredenhagen [see 11]
and by Haag [1, page 180].

The geometrical approach based on functional integrals and formal (Verma
modula)
representation theory is in no way limited by positivity requirements since it
only explores free-field-like (integrable) system. In physical ``real life"
the von Neumann algebra methods of algebraic QFT are however indispensable.

In contrast to the functional integral approach, the locality principles in
algebraic QFT
is taken care of in every step of the arguments viz. the title and the
content of a recent book [1].

 In this paper we will only use
regular states on nets and then topological data will only appear in suitably
defined intertwiner subalgebras (centralizer algebras) which are of the $II_1$
Jones type and contain mapping class group matrices. The numerical invariants
of 3-manifolds are from our viewpoint  computable by suitable
limits in the spirit of vacuum polarization effects, which amounts to the same
as ``$\rho_{reg.}$" in [7].

If one adds spinor-matter to 3-d pure Chern-Simons actions, one expects a
remaining
genuine III$_1$
net  as the ``gauge invariant" physical part of the model [12]. In fact one
expects the principle role of the formal local gauge principle in
low-dimensional
QFT's to be just this: a method to find interesting observable (plektonic)
subalgebras inside
the well-known algebras of Bosons and Fermions (but unfortunately not for the
construction of the very non-classical and yet unknown
field algebra). In the 4-d gauge theories, there is
of course the additional problem of understanding the true quantum nature of
``Magnetic" fields and ``Stokes theorem" (on the same conceptual level as
the notion of charges in the theory of superselection sectors) and the deeper
meaning of ``quark confinement".

The present picture of 4-d QFT from the algebraic viewpoint is less rich then
that
of low-dimensional QFT. The reason for this is that the standard spectral gap
assumption for the energy-momentum spectrum leads to the Buchholz-Fredenhagen
semiinfinite space-like cone localization (the cones can be made arbitrarily
thin)
[63], which in turn yields permutation group statistics and compact group
symmetry.
The situation would change dramatically if one finds a physical argument in
favor
of infinite energy objects which allow no better localization than that around
a
core being a space-like hypersurface. The quark idea (outside the
short-distance
regime) suggests that it may be advantagous to think about quantum states
(i.e. no gauge artifacts) with infinite energy. Quarks as Wigner-states with a
finite mass and spin $1\over 2$ are too naive and not very well digestable by
algebraic
QFT. With the exception of this short interlude the paper will be focussed on
low-dimensional QFT only.

While at the issue of physical interpretation, we find it worthwhile to mention
the
difficulties, which algebraic QFT encounters with the notion of chiral
conformal
QFT ``on Riemann surfaces". In the special case of genus $g=1$, the algebraic
approach (as well as the Wightman approach based on the use of point-like
covariant fields)
yields two nets: one ``living" (in the sense of localizability) on the
Minkowskian
cycle and the other (beeing also non-commutative and lacking a Feynman-Kac
representation)
on the Euclidean cycle. Between these two cycles there exists (as in standard
QFT's)
the analogue of the Bargman-Hall-Wightman region [13] of analyticity of
correlation
functions,\footnote{$\ \ g=1$ belongs to the $L_o$-Gibbs state correlation
functions.
A conceptually acceptable desciption of $g\ge 2$ in terms of the notion of
algebra
and states is presently not available although some formal observations on
avering conformal
correlations by Fuchsian groups exist [64].}
which, at least a priori, has nothing to do with the localizability implicitely
 contained in the word ``living on...". For this reason, it is the more
surprising that we find the presence of mapping class matrices belonging to all
geni,
even if we presently do not know the physical interpretation of this finding.
How do nets on $S^1$ know about mapping class groups for all geni?
Does this observation mean that plektonic physics is better prepared (than
fermionic
or bosonic) to accept ideas about quantum gravity? Is it related  to the
absence
of degrees of freedom of gauge- and metric-fields  in a quantization approach
to
low-dimensional QFT's?

Concerning the aforementioned first problem of anyonic or plektonic free fields
one has an extremely
useful physical picture [14], as yet without sufficiently general explicit
constructions. The picture
emerges from a combination of the old Coleman-Mandula [8] ideas with the
results
from the bootstrap program of introducing 2-d integrable models via factorizing
S-matrices and formfactors of the would be fields [15].

In S-matrix
language the important observation (for our purpose) may be phrased
in the following manner. If, in a 4-d local QFT one performs the ``complete
cluster limit" on the S-matrix i.e. the limit in which all the centers of
individual
particle wave packets become infinitly space-like separated, then all
scattering
effects, (first the inelastic ones and then also the particle conserving
processes)
die out and only $S=1$ remains in the limit.

A trivial S-matrix however is believed
to belong to the local equivalence class (Borchers class) of a free field,
although
a mathematical proof exists only under special additional conditions [17].
The complete clustering idea of S is used here to produce the ``freest"
permutation group statistics QFT which turn out to be identical with a free
field theory in the usual technical sense (i.e. the higher point functions are
products of
two-point function, with the appropriate Wick-combinatories).

Let us now jump from
4 down to 2 space-time dimensions. In that case the same extreme cluster limit
of the
S-matrix is also expected to lead to an asymptotic vanishing of all inelastic
processes and
part of the elastic ones. Only the elastic two particle scattering
$S^{(2)}_{limit}(\theta)$
 $(\theta$=rapidity) should survive\footnote{The higher elastic processes have
a
smoother threshold behaviour leading to a faster decrease in the spatial
separation of wave packets.}
 and then all the higher elastic processes for consistency
reason have to go through 2-particle scattering.

 $S^{(2)}_{limit} (\theta)$
allows no further asymptotic space-time simplification since a separation of an
interaction T-part from a noninteracting  part in the decomposition $S=1+iT$ is
physically meaningless
 (both parts contain the same momentum space $\delta$-functions).

The Yang Baxter equation emerges as a consistency condition from this picture.
Therefore
this picture is ``quantum integrability" par exellence without recourse to a
classical
parallelism, conserved currents etc. The nontrivial part of this picture is of
course the statement that the limiting S-matrix belongs to a fullfledged
localizable
QFT in its own right (partially obtained via the so-called form-factor program
[18]).

So quantum integrability is conceptually much easier than classical
integrability,
which in the spirit of Bohr's correspondence principle should be derivable in
an
appropriate semiclassical approximation (not an easy matter!).

The particles to which the limiting S-matrix and its affiliated localizable QFT
belong to  plektons in special cases (i.e. the internal
structure of S may be complicated than that describable by symmetry groups)
but generically they are ``kinks" (which are not DHR-localizable but rather
belong to
half space localized solitons) for which
statistics notions are not applicable (statistical ``schizons", whose operator
commutation relations can be changed by neutral operators i.e. without changing
charges [19]).
In the calculational scheme of Smirnov [18] this last observation is turned
into
a virtue by selecting for each interpolating field a bosonic representative for
which the formfactor computations seem to be easier.

Now, finally, let us apply the picture
to 3-d QFT. In that case we know that the generically admissable statistics is
braid group statistics. On the other hand the complete cluster limit does not
allow
for an energy dependent limiting S-matrix as it was the case in d=4. This
leaves
only the possibility of a piecewise constant S-matrix (in momentum space) with
discontinuous jumps if asymptotic directions (directions of momenta) go through
degenerate situations where at least two direction become parallel.
The properly defined cross-section for such a S-matrix would vanish. The values
in
the different nondegenerate components are given in terms of constant
braid-R-matrices.
This picture gets additional support by looking at the structure of scattering
states. [20]

Insufficient knowledge of low energy analytic momentum space properties
in the presence of braid group statistics has up to now prevented a proof a
l\`a
 Coleman-Mandula which is based on the usual (permutation group) statistics
analytic
properties of dispersion theory.

An explicite construction of anyonic or plektonic
free fields (i.e. their particle form-factors) belonging to that limiting
piecewiese
constant S-matrix is presently even more
remote. Without doubt such objects, if they exist, should be considered as the
3-d version
of quantum integrability (i.e. if those piecewiese constant S-matrices admit
localized
interpolating fields).

Note that such a picture leads to a new class
division of theories according to their long distance properties: each
interacting
theory, as complicated is it may be, has precisely one integrable companion
namely that belonging
to $S_{limit}$. Up to now one only knows a short-distance universality
principle
formulated in terms of ``conformal companions" of quantum field theoretical
models.

The second problem, namely (internal) symmetry is closely related to the
previous one.
In order to see this, I remind the reader of a well-understood (but
unfortunately not
so well-known) fact that in addition to the standard description of QFT with
permutation
group statistics in terms of local boson or fermion fields with tensorial
compact
group transformation, there exist a physically equivalent (regarding local
observables
and the scattering matrix of charged particles) formalism based on nonlocal
para-statistics
fields (the exchange-algebra fields for the permutation group) which have
no group transformation properties and fulfill R-matix commutation relations
with
$R^2=1$ [43].

If one wants to prove e.g. that the free field equation (considered as a
defining 2-sided ideal)
or $S=1$ leads to the analytically correct correlation functions of a free
field
theory, it would not be advisable to use para-fields. But with local tensorial
fields,
the result follows easily  from the field equation and the well-known analytic
properties of correlation functions which follow from causality and spectral
properites
of tensoriel fields. \footnote{By a trivial extension of the Jost-Schroer
theorem [13].}

The analytic structure od $d=2,3$ exchange algebra
fields is much more complicated than that of para-fermions or bosons.

So if one would
find a symmetry concept which could replace the tensorial formalism in the
presence
of braid group statistics, then this  could be of great practical value.
Nonrelativistic plektons and their (non understood) ``quantum" symmetries could
allow to obtain a new symmetry breaking which may be responsible for the
fractional Hall effect and High-$T_c$ superconductivity. Braid group
statistics,
even at its present state of poor physical understanding has already produced
an extremely rich
collection of numbers: rational statistical phases (related to electromagnetic
properties?)
and algebraic integers in the form of statistical dimensions
(related to amplification factors in cross section or
thermodynamical properties?) which are generalized Casimir factors counting
degrees
of freedom.\footnote{Statistics of (quasi-)particles and the ensuing change of
composition laws of spins and charges is the only difference
between 4-d and low-dimensional QFT. This is expected to lead to significant
differences of plektonic phases from the Fermi-liquid phases in condensed
matter physics. The problem of classifying these phases is similar (but in its
analytic aspects more difficult) to the classification of 2-d conformal QFT's.}
But one lacks a good understanding of their localized field carriers, which
would be necessary to work out experimental consequences.

Present attempts using quantum group ideas  do not seem to go into this
direction.
They often contain artificial integer numbers (the representation
dimensionality of
the $q$-deformed group), treat the notion of conjugates (antiparticles) in an
extremely
unsymmetric way, and generally lead to non-unique systems of fields not
relatable
by Klein-Transformations [22].

There are rather convincing arguments that the quantum
symmetry aspect should be discussed together with space-time symmetries
including
the TCP symmetry and also with modular Tomita-Takesaki theory and not as an
isolated
kinematical (combinatorical) property beeing apart from the rest of QFT.

In the following
I will present some recent results on symmetries obtained by modular theory
(in most parts within the framework
of algebraic 2-d conformal QFT) and forget the presently inaccessable problem
of
free plektons. But it should be clear that my motivation (for presenting these
partial results on symmetry) is entirely physical.

The paper is organized as follows.

I start in section II with a net indexed by intervalls on a line and build up a
conformal
net on the circle. The modular Tomita-Takesaki structure, which is heavily used
in this
construction, also leads to a Euclidean net whose construction is only scetched
very briefly.

Section III recalls the derivation of the universal observable algebra with its
nontrivial center. It contains an ``invariant symmetry algebra" which lends
itself
to a generalization of the Kac-Wakimoto [23] formula.

In section IV we recall the proof of the TCP theorem in the framework of the
exchange
algebra. In the case of the ${\bf Z}_{2N}$-anyon algebra we show the
coalescense of
this theorem with the standard Tomita-Takesaki modular theory on the
half-circles
and argue that this state of affairs prevails for all models with vanishing
``defect projector" which measures the deviation of statistical dimensions from
integers.

In section V we study the ``selfduality", observed globally as the symmetry of
Verlinde's matrix S, on the local net level. Here another tool of modular
theory
is used: Longo's ``canonical endomorphism". Again we employ the ${\bf Z}_{2N}$
anyonic
model as an illustration of perfect selfduality.

Finally in the last section we use an old, but still mysterious observation of
Vaughn Jones
in order to make some speculative remarks on the possible nature of the
non-invariant
aspect (i.e. beyond the symmetry algebra of section III) of a non-commutative
link between external and internal symmetry.
This section belongs to the realm of fantasy and only serves the purpose to
open
the eyes for radical new possibilities of field algebras.
   done by space-time regions in Minkowski-space, but by intervals
   on a line (a right or left light ray of a 2-d.\ conformal QFT).
   We assume a translation-invariant vacuum state $\omega _0$ and
   construct via the GNS construction the vacuum-representation
   $\pi _0({\cal A})$ and the affiliated
   von Neumann algebra $M = \pi _0({\cal A})''$ as well as
     the unitary positive
   energy translation $U(a)$ implementing the assumed
   translation automorphism and leaving the GNS vacuum vector
   $\Omega $ invariant.

We consider the half-line algebra $M_+ =
   \pi _0({\cal A}(0, {\infty} ))''$ on which the right translation
   $U(a), a>0$ acts like a one-sided compression.
   Under these circumstances Borchers proved the following
   theorem [24]:\\[3mm]
   {\bf Theorem(Borchers):}~~ {\em  The modular reflection $J$ and the modular
      operator $\Delta ^{it}$ of the pair $(M_+,\Omega )$ have
      the following commutation relation with $U(a)$:
\begin{eqnarray}
  JU(a) J & = & U(-a) \\
  \Delta ^{it} U(a) \Delta ^{-it} & = & U(e^{-2\pi t}a),
               ~~~t, a \in R\nonumber
\end{eqnarray} }

It was already known before, that the abstract Tomita-Takesaki
modular theory (which affiliates a reflection $J$ and a one
parameter group $\Delta ^{it}$ with a von Neumann algebra and a
cyclic and separating state vector) has a deep physical and
geometric significance in ordinary Wightman QFT if one looks
at algebras belonging to specific regions (``wedges''):
the Tomita reflection $J$ becomes (up to a $\pi $-rotation)
the TCP operator related to the appearance of antiparticles
and $\Delta ^{it}$ a special $L$-subgroup leaving that
wedge region invariant. The operator $\Delta ^{it}$ in Borchers theorem
 is clearly the dilation.

{}From the translations and $J$ one can build up a new net
 (as Borchers did)
 by
starting with $M_+$ which is by construction covariant under
the M\"obius subgroup leaving ${\infty} $ invariant.
If the new net has the self-reproducing property, i.e.,
 is equal to the original net, the original net
 is admissable for the subsequent consideration. If, however,
  the original net is perverse and fails to have this
  property, we work with the new net
  (which is canonically affiliated with the old one
  and has the self-reproducing property by
  construction) and forget the old one.

For the full M\"obius-group including the change of ${\infty} $
one needs the following ``quarter-circle assumption''
(we use $S^1$ instead of $R \cup \{{\infty} \}$):\\[3mm]
 {\bf Assumption:} {\em Essential duality holds for the interval $\left( -
      \frac{\pi }{2}, \frac{\pi }{2}\right) $:}
      \\
      i.e.,
\begin{eqnarray}
&&    {\cal L} \left( -\frac{\pi }{2},\frac{\pi }{2}\right)
                = {\cal L} \left( \left( -\frac{\pi }{2},
                    \frac{\pi }{2}\right) '\right) '\\
&&      \mbox{with}~~ {\cal L} (I): = M(I')', \nonumber
\end{eqnarray}
and the modular conjugation $\tilde{J}$ of ${\cal L}$ permutes
 the quarter circles, which one obtains by intersection
 the old half circle with the new one, so that we obtain
\begin{equation}
 \tilde{J} M_+ \tilde{J} = M_+.
 \end{equation}
 Then one finds [25]:\\[3mm]
 {\bf Theorem:}~~{\em The group generated by $U(a), \Delta ^{it}$ and
       $J \cdot \tilde{J}$
is
the full M\"obiusgroup and the net generated from the four quadrant
algebras is M\"obius covariant.}\\

 Again we dismiss the input
 net if it is not identical to this final fully covariant net.

This theorem has been recently derived from a much wider and natural
 mathematical scheme than in [25] called ``half-side modular inclusion''
 by H.W.-Wiesbrock~[26].

  Note that essential duality is
  weaker than Haag duality. It also holds for nets
   generated from the energy momentum tensor for $c_{virasoro} > 1$
   where, according to Buchholz and Schulz-Mirbach [27], Haag
   duality breaks down for the net on {\bf R} generated by the enery-momentum
tensor.

 The above theorem suggests that the
   B.-S.M. mechanism of violating Haag duality and full conformal
   invariance by ``drilling a hole'' into $S^1$ (i.e., going
   from $S^1$ to $R$) is the only one, by which translation
    covariant nets indexed by intervals on the line $R$ can fail
    to be a chiral conformal field theories. But then there always
    exists a canonically affiliated M\"obius covariant $S^1$-net.
    For such a covariant net one can directly
     show that the stronger Haag duality hold for each interval [28,29].

Replacing the assumed geometric translation by the product of two abstract
Tomita-reflections one could even remove the last vestiges of geometry from
the quantum-physical assumptions [26].

Since the T.-T.~modular theory was shown to build up
M\"obius-invariance from pure algebraic inclusion data (indexed by
 intervals on $S^1$), it is tempting to ask whether diffeomorphisms
  beyond the M\"obius transformation can also be generated from algebraic
  net inclusions and suitably chosen states.

Let us look at the simplest such transformation (which
  together with the M\"obius transformation generates the
  diffeomorphism group):
\begin{equation}
  Dil (t)^{(2)} z:= \sqrt{ Dil(t) z^2}.
  \end{equation}
  Here we use the complex coordinate $z$ for the circle,
  $Dil(t)$ is the usual dilation, i.e., the modular
  group of the pair $({\cal A}(S_+), \Omega )$ and the
  resulting $Dil^{(2)}$ denotes the ``dilation'' in the ``second
  Virasoro sheet'', i.e., a geometric transformation
  built from infinitesimal generators $L_{\pm2}, L_0$ which
  ``dilates'' with 4 fixed points: $z=1, i; -1-i$ instead of
  the $\pm 1$ fixed points of $Dil$.

This suggest to consider
  a doubly localized algebra
\begin{equation}
  {\cal A}(S_1) \vee {\cal A}(S_3)
  \end{equation}
  where $S_1$ and $S_3$ are the first and third quadrants of
  the circle (counterclockwise).

 The corresponding state vector
  candidate for a modular pair cannot be the vacuum state-vector
  $\Omega $.
  Here the split property [30] gives us clue. In conformal field theory
  this property follows from a controll of the asymptotic
  spectral density of the rigid rotation operator $L_0$ [28],
  and holds in all known conformal models [31].
  It has the consequence that algebras like the previous one
  are spatially isomorphic to tensor products.
  \begin{equation}
  {\cal A}(S_1) \vee {\cal A}(S_3) \sim {\cal A} (S_1) \otimes {\cal A}(S_3)
\sim {\cal A}_I (S_+)\otimes {\cal A}_{II}(S_+).
  \end{equation}
where ${\cal A}_{I,II}$ are two copies of the halfcircle algebra.

It is well-known that tensor-product algebras together
with tensor product state vectors have also a tensor product
modular operator. But our state vector is prevented to have this form
as a result of the presence of a unitary equivalence
transformation.
We define the state on the doubly localized algebra by the vacuum product state
on the single components. The use of the ``natural cone" of the doubly
localized
algebra allows for a unique representation of that state in terms of a state
vector $\phi$. In order to understand the modular theory for this situation we
investigate the KMS property of
\begin{equation}
(\phi, A_I A_{II} B_I B_{II}\phi) := (\Omega,\hat A_I\hat B_I\Omega)(\Omega,
\hat A_{II}\hat B_{II}\Omega)
\end{equation}
with: $A_I,B_I\epsilon {\cal A}(S_1),\quad A_{II}, B_{II} \epsilon{\cal
A}(S_3),\quad
\hat A_I,\hat B_I\epsilon {\cal A}_I(S_+),\quad \hat A_{II},\hat B_{II}
\epsilon
{\cal A}_{II}(S_+)$

The modular operator $\triangle$ appearing in the KMS relation on the left hand
side:
\begin{equation}
(\phi,A_I A_{II} B_I B_{II} \phi) =(\phi,\triangle B_I B_{II} \triangle^{-1}
A_I A_{II} \phi)
\end{equation}
can easily be expressed in terms of the U-transformed modular operator
$\hat\triangle \otimes \hat\triangle$ belonging to the right hand side where
($U$ is the unitary equivalence transformation).
\begin{equation}
\phi=U(\Omega\otimes \Omega)
\end{equation}

The problem of finding the correct state vector has therefore
been reduced to the question of whether one can construct a
unitary operator with a geometric interpretation.
Imagine that we start with the two copies
 on $S_+$ i.e. $\hat{\cal A}_{I,II}(S_+)$ and the vacuum state vector $\Omega$.
These algebras are certainly isomorphic to ${\cal A} (S_1)$ resp. ${\cal
A}(S_3)$.
 Then  the covering transformation $z\rightarrow z^2$ maps $S_1$ and $S_3$ onto
$S_+$. As an automorphism of ${\cal A}(S)$
this is meaningless; it is not a diffeomorphism. However,
restricted to the doubly localized algebras it makes sense.
So, there could exist a ``partial automorphism''
(i.e., not extendable by the geometric formula to ${\cal A}(S^1)$)
of the doubly localized von Neumann algebra which can be implemented in
the vacuum representation by a unitary operator $U_{cov.}$ transforming
from the tensor-product space to the original
Hilbert-space, such that (12) with $U^{-1}=U_{cov.}$
is the  state vector yielding $\triangle^{it}= Dil^{(2)}(t)$
as a modular object. Conditions for the uniqueness of the implementing unitary
are known by the ``natural cone" construction [1, Thm. 2.2.4]

The only piece which is lacking in order to
have a proof, is the existence of the partial automorphism entailing that of
$U_{cov.}$ with the
geometric action on fields. Unfortunately we have not been able to find
a structural analytic argument similar to [27]. However in models
which have a ``first" quantization like the Weyl algebra one has a controll
of the partial automorphism.
 We will treat this problem in a future publication.

Note that this method which is based on the use of the KMS property only deals
with the modular group and does not determine the Tomita reflection and  its
possible geometric manifestation in the form of Haag duality.

The rigorous clarification of the general situation would be important because
it allows
for a modular (i.e., pure quantum physical)
understanding of the Virasoro charge, the origin of
diffeomorphisms and the intrinsic
meaning of energy-momentum tensors in the framework of the
algebraic net theory.

     The way in which the Tomita-Takesaki modular theory converts
     formal net indexing by intervals into geometry and physics is really
     surprising. There is more to come later on.

For later use we also explain schematically the construction fo Euclidean
chiral theories. The starting point is the observation,
that the existence of a positive natural cone
(Araki, Connes [32]) in the Tomita-Takasaki modular theory
entails the existence of another scalar product between state
vectors of the form\footnote{The inner product is the same as
that which appears in the Connes bimodule theory if one restricts
 the bimodule to the diagonal action.}
\begin{equation}
  \psi_A  = \Delta ^{1/4} A \Omega
\end{equation}
\begin{equation}
\langle A,B\rangle := (\psi_A,\psi_B) = (JA^*J\Omega,B\Omega)
\end{equation}
In our field-theoretic application $A \in M_+$, $\Delta ^{it}$
is the dilation and $\Omega $ is the  (with respect to $M_+$)
cyclic and separating vacuum state vector.  This yields another state-space
${\cal H}_{eucl.}$ with a different involutive and positive
star-operation on which the old translation in positive direction
becomes a contraction operator on ${\cal H}_{eucl}$.
 This in turn allows for a unitary analytic continuation
 which should be thought of as a translation on the imaginary
 axis. Using this euclidean translation, one can build
 up another net, whose localization intervals are to be placed
 on the imaginary axis and which is very non-local relative to the
 original real-time net. The only common elements are the non-local
 analytic (with respect to the modular operator $\Delta ^{it}$)
 elements.

We call this euclidean theory the cartesian euclidean theory.
 A richer and more interesting euclidean theory is the radial
 euclidean theory. Its construction also starts
 from the upper half circle, but instead of the one-sided
 translation one uses a two-sided compression defined in
 terms of the modular dilation $\tilde\Delta^{it} $ of the
 right-hand half circle. In this case the analytic continuation
 of the two-sided contraction yields the euclidean rigid
 rotation, i.e., a rotation whose localization properties should
 be pictured in terms of a radial periodicized coordinate.
 It is remarkable that the Tomita reflection of the euclidean
 theory is identical to the original star-operation which now
 acts like a charge conjugation i.e. as a
 bounded operator on ${\cal H}_{eucl.}$ (and in turn the euclidean-star
  becomes related to the real-time Tomita reflection).
 \section{Superselection Sectors and the Invariant Symmetry Algebra.}
 For a net indexed by intervals on a circle (rather then on a line),
 the previously constructed vacuum representation does
 not admit sectors, i.e., one cannot find interesting
 endomorphism [7]. On the line $R$ one could find such endomorphism
 on the quasilocal $C^*$ algebra ${\cal A}_{quasi}$ (the DHR algebra) which is
directed
 towards infinity [6], but then one would have wrecked conformal
 invariance.

It turns out that the correct observable algebra on
 $S^1$ which does not have these faults can be obtained by
 a universal construction: it is the free $C^*$-algebra
 generated from the vacuum-represented net amalgamated over all local
 relations [34]. Using the information one has from superselection theory,
 one finds that this universal algebra ${\cal A}_{univ.\ }$
 has  global self-intertwiners which lead to a nontrivial
  center generated by invariant (abelian) charges [7].

 As the old (DHR)
   quasilocal algebra ${\cal A}_{quasi}$, it is
    uniquely determined from the
   net data in the vacuum representation. The universal
   observable algebra ${\cal A}_{univ.}$ also has an analog representation
   theory in terms of localized endomorphism as the DHR
   theory, only the meaning of ``localized'' and ``transportable''
   is somewhat more restricted [7].

The
   derivation of the braid group structure is also similar to the standard
   theory [7]; there is, however, one additional item as already mentioned:
   the appearance of a global self-intertwiner
   $V_\rho$ which ``flips'' the statistics
   operator $\epsilon (\rho ,\sigma )$
   to $\epsilon ^*(\sigma ,\rho )$.

Together with the change
   of localization this fact accounts for the different
   properties of the exchange algebra, which, as in the standard
   theory, is constructed in a canonical fashion from the ${\cal A}_{vac.}$
   data together with the endomorphisms and constitutes a kind of
   ``nonlocal'' extension of ${\cal A}_{univ.}$.
   \\[3mm]
   {\bf Theorem (FRSII [7]):}~~{\em In theories with a finite
   number of irreducible endomorphism
    (finite depth in the Jones theory, rational theory in conformal
    QFT) the exchange algebra is localizable on an N-fold covering
    (related to the depth) of $S^1$ and its $R$-matrices represent
    the braid group of the cylinder (generated by applying
    endomorphisms to the generators $\epsilon (\rho ,\rho )$
    and $V_\rho $).}\\

In this formulation we adapted the QFT to chirally conformal
    field theory, a generalization to the $3-d$ case is
    straightforward [7].

 In the framework of the DHR
    theory  the exchange algebra (see also the beginning of the next section)
    is the reduced ``field bundle'' [5] (a bimodule
    over ${\cal A}$ with a $C^*$~algebra structure),
    whereas in the Wightman-framework it was first abstracted from
    nontrivial conformally invariant models [35].

It is profitable to define a subalgebra ${\cal G}^{inv.}$
 of ${\cal A}_{univ.}$
generated by the invariant charges (which depend only on equivalence
classes of endomorphism), and their images
 under application of $\rho$'s. We do this as follows.
We first pick one endomorphism $\rho _i$ for each sector (the
 ``rational'' situation being always assumed) where all
 $\rho _i$'s are localized in one proper intervall
 on $S^1$. The theory then assures the existence of localized
 isometric intertwiners $V_i$ such that
\begin{eqnarray}
&&   \rho : = \sum_{i} V_i\rho _iV_i^* \\
&&     \mbox{with}~~~ \rho ({\cal A}_{univ.})
      \subset {\cal A}_{univ.} ~~~\mbox{and}~~~ V_i\in
       {\cal A}_{univ.} \nonumber
\end{eqnarray}
The centralizer algebra
\begin{equation}
M = \lim \mathop{\cup}_{n} \left( \rho ^n, \rho ^n\right)
\end{equation}
with $(\rho ^n,\rho ^n)$ = space of intertwiners
$\rho ^n\rightarrow \rho ^n$,
is the inductive limit of finite matrix algebras.

 $M$ allows for a
restriction to the braid-group algebra $CB_{\infty} $, whereupon
the tracial state on $M$ (obtained by the field-theoretic
left inverse of $\rho$) aquires the Markov property.

 The
above reducible $\rho $ and the intertwiner-spaces appear in
the work of Rehren [40]. Inductive limits of this kind are
known to be the hyperfinite II$_1$ factor $R$, following the well-known
arguments based on ``commuting squares" [55].

Let us now consider the subalgebra generated by elements
of the form
\begin{equation}
\rho (C_{i_1} \rho (C_{i_2}) \dots \rho (C_{i_n})\dots )
\end{equation}
with $C_i, i=1 \dots N$, the invariant global charges
of ${\cal A}_{univ.}$ Clearly,
these  operators are in $(\rho ^n, \rho ^n)$, and their
 representation in terms of the path-space of intertwiners
 [37] shows that they are diagonal, in fact they are linear combinations of
 minimal projectors.
 The general intertwiners in $(\rho ^n, \rho ^n)$ are
 ``strings'' [6], i.e., pairs of intertwiner-path
  with equal length  which correspond
 to matrix units. Clearly, these diagonal elements form a subalgebra
inside the centralizer algebra.

In the following discussions we will call this subalgebra (for obvious
reasons)\footnote{Strictly speaking only the GNS representation
of ${\cal G}^{inv.}$ with class invariant states (Gibbs
or tracial states only depend on the sector $[\rho ]$) deserve this name.}
 the invariant symmetry algebra ${\cal G}^{inv.}$.  Its lowest nontrivial
 element (again restricting to the rational nondegenerate situation
 [36]) evaluated in the vacuum representation is the numerical
 matrix $S$[7].
 \begin{equation}
 \pi _0 \rho _i (C_j) = S_{ij}.
 \end{equation}
 This matrix, (together with a diagonal matrix $T$) first abstracted from
characters of chiral
  conformal models by Cardy, Capelli, Itzykson and Zuber, then related
  to fusion rules by Verlinde [38].
  and generalized to a wider geometrical picture
  of chiral conformal field theory by Moore and Seiberg [39],
  was finally shown to relate fusion rules and
  monodromy properties (statistics character) independent
  of conformal QFT by Rehren [40]. By the present work we may now add the
remark
that the universal algebra on the circle (as well as the 3-d universal algebra)
contains mapping class group representations for all geni.

Let us look now at more
  general elements of ${\cal G}^{inv}$
  evaluated in the vacuum representation. These are
  not just numerical-valued, since the multiple application
  of $\rho $ lead inside the non-central parts of the original
  algebra ${\cal A}_{univ}$ (except for the
  abelian case).
In order to get to numerical quantities, one can use the tracial state.
It leads to numerical matrices which generalize $S$ and are
made up from $S$ and traces on minimal projectors.
These multi-indexed matrices will be shortly identified with the known
mapping class matrices [41].

 On the
  other hand ${\cal A}_{univ.}$, and therefore ${\cal G}^{inv.}$,
  allows for a Gibbs temperature-state in the vacuum sector
  which is quasiequivalent to the vacuum state.
  Formally this state becomes (up to a possibly infinite
  normalization factor) tracial in the infinite temperature
  limit and therefore, restricted to $M$ and ${\cal G}^{inv.}$ is proportional
  to the Markov tracial state (which the left inverse
  produces on the right-hand centralizer algebra.)
  with a possibly divergent normalization factor.

 This observation
  explains why inclusion data like Jones indices, which are
  to be found in the centralizer algebra, can also be expressed
  (\`a la Kac Wakimoto [23]) in terms of infinite temperature
  limits of characters (partition function) of ${\cal G}^{inv.} \subset
  {\cal A}_{univ.}$.

The most surprising fact is that the tracial state on ${\cal G}^{inv.}$
leads to (as a generalization of the matrix $S$)  multiindexed
matrices which represent higher genus Riemann surfaces mapping
class group.
This is seen most easily from the known graphical representations of
these matrices [41]. The crossings in those pictures correspond to the
 matrices $S$ and the trilinear vertices to the intertwiners.
They occur always in pairs and these pairs correspond to
projectors whose graphical representation is of the form of parallel path.

 This observation on the
presence of mapping class matrices for all geni in a chiral conformal
field theory on a circle is very startling.
It should be take as a hint of a new yet unknown quantum
 symmetry of which we presently only see its invariant part.

Note also that each operator in ${\cal G}^{inv.}$ (and M) in the
temperature state has a (generalized) Kac-Wakimoto representation
in terms of an infinite temperature limit.
This is so since ${\cal G}^{inv.}$ permits two types of states:
the temperature-states coming from the physics of ${\cal A}_{univ.}$
restricted to ${\cal G}^{inv.}$, and the more algebraically motivated
states on the intertwiner spaces defined by the iteration of
left-inverses [5,6]. They also appear in topological field theory [10],
if one interprets a functional integral together with the rules
 for its calculation as the assigment of a state to a
 global algebra which is similar to $M$, i.e., lacks the
 localization property of ${\cal A}_{univ.}$. What
 is the geometric meaning (if any)
 of the class invariant (referring
 to the dependence on individual
 endomorphisms)
  thermal expectation values of operators in ${\cal G}^{inv}$
 before one takes the infinite temperature limit?
 I do not know the answer.

\section{The TCP Theorem of the Exchange Algebra and its
          Modular Interpretation for Anyons.}
 We first recall our findings [7] on the TCP operator
 of the exchange algebra. The latter is generated
  by operators $F$ which act on the Hilbert-space
  consisting of $N$ copies of the vacuum space:
  \begin{equation}
  {\cal H} = \mathop{\oplus}_{\alpha } (\alpha ,{\cal H}_0)
  \end{equation}
  according to:
  \begin{equation}
  F(e,A) (\alpha ,{\cal H}_0) = (\beta ,\pi _0
              (T^*_e\rho _\alpha (A))\psi )
\end{equation}
  with $A \in {\cal A}_{univ.}, T_e \in {\cal A}_{univ.}:
        \rho _\beta  \rightarrow \rho _\alpha \cdot \rho $
        and $\rho _\alpha ,\rho _\beta ,\rho $ taken from
        an arbitrary but fixed finite set of reference-endomorphisms.

If $e$ runs through the finite set of all ``edges'' $e$
of the form ($\rho _\alpha ,\rho ,\rho _\beta )$
and $A$ through ${\cal A}_{univ.}$, then the $F$ generate a
$C^*$ algebra (exchange algebra or reduced field bundle)
which, since it is an operator algebra in a concrete $H$-space,
also allows for the von Neumann closure.

 In contrast to Boson-(CCR)
or Fermion-(CAR) algebras, the charge conjugation in this algebra
differs significantly from
the $*$-operation.

The $*$-operation has the
form:
\begin{equation}
F(e,A)^* = \frac{d_\rho }{\chi_\rho } \sum_{e^*}
    \eta_{ee^*} F(e^*,\bar{\rho }(A^*) R_\rho )
 \end{equation}
     where $e^*$ is of the adjoint form $(\beta ,\bar{\rho },\alpha )$
     and the coefficients ($d_\rho $ statistical, dimension, $\chi_\rho $
    a phase and  $\eta$ a structure constant matrix) are as
     in [7].

On the other hand the antilinear charge conjugation
     of the exchange algebra reads as:
     \begin{equation}
     \overline{F(e,A)}: = \sqrt{ \frac{d_\alpha }{d_\beta }}
                  \frac{d_\rho }{\chi_\rho } \sum_{\bar{e}} \zeta_{e\bar{e}}
                  F (\bar{e},\bar{\rho }(A^*)R_\rho )
\end{equation}
     with $\bar{e}$ of the form $(\bar{\alpha },\bar{\rho },\bar{\beta })$
            (and $\zeta$ another structure matrix). It fulfils
            \begin{equation}
    \overline{\bar{F}} = \frac{\chi_\beta }{\chi_\alpha } \cdot F
 \end{equation}
    where
    the phase can be adjusted to be $\pm 1$ ($+1$ in the non-selfconjugate
    case).

In analogy to the standard case, one may define an operator:
\begin{equation}
S_\pm: F(e,A) \Omega  \rightarrow \overline{F(e,A)} \Omega ,~~~~~
          F \in {\cal F}_{red}(R_\pm)
 \end{equation}
    which together with its adjoint turns out to be densely defined, hence
    closable and therefore permits a polar decomposition [9]. Some
    simple
    considerations
    using ``coordinates'' of point-like fields [7]
reveal that
    this polar decomposition is geometric
    \begin{equation}
    S_\pm = \kappa ^{\pm\frac{1}{2}} \theta V(\pm i \pi )
  \end{equation}
    where $\kappa $ is a statistics phase, $\theta$ the TCT operator
     and $V(\pm i\pi )$ the analytic continuation of the dilation on
     ${\cal H}$.

 This geometric result parallels the previously
     obtained result based on the use of the Tomita-Takesaki
     modular theory for the observable algebra [27].

     However, these formulae in the
    in the case of our exchange algebra  cannot be backed up
      by Tomita-Takesaki modular theory, since this theory
      requires the (positive-definite) $*$-operation
      (and cannot be done with the involutive antilinear charge-conjugation)
in formula (19).

 In our paper [7] we speculated
      about the possible existence of a ``twisted'' Tomita-Takesaki theory
      relating a twisted involution with a twisted commutator.

But
      perhaps a more conservative idea is to follow the mechanism by which
      the ``would be modular theory''
      in the case of trivial monodromy $\epsilon ^2 = 1$, (i.e.,
      the permutation group exchange algebra) can
 be converted into a true modular theory.

 In that case,
      we know, according to Doplicher and Roberts [42] that there
      exists a genuine field algebra of Bosons and Fermions
      from which the exchange algebra can be obtained by removing
      multiplicities [43], i.e., making the Hilbert-space smaller.

      The prize one pays for converting the Bosons and Fermions
      with a non-abelian symmetry group into Para-Bosons and
      -Fermions (on which no nonabelian compact group can act
      since the $H$-space lacks the necessary multiplicities), is
      a certain amount of non-locality, now expressed in terms of $R$-matrix
      commutation  relation ($R^2 = 1$) instead of local
Boson or Fermion
      (anti)commutation relations for space-like distances [43].

      In fact, one could contemplate that the insistence in a
      geometric (local) version of the standard Tomita-Takesaki
      modular theory would determine the smallest enlargement
      of the $H$-space of the permutation statistics
      exchange algebra and in this way could
   yield a direct operator-algebraic
      proof of the D.R. theorem (which, in its
      present form contains a significant amount of
      $C^*$category theory).

In any case, the previous considerations
      strongly suggest that the search for a new ``quantum symmetry''
      behind braid group statistics should be persued in
      conjunction with the modular Tomita-Takesaki
      theory and the TCP properties of QFT.

It is interesting
      to note that the special abelian case of $Z_{2N}$ anyons can
      still be covered by the standard Tomita-Takesaki theory.
      As in the case of Fermions [3], one has to introduce a
      suitably ``twisted''or ``quasi"- commutant (and no ``twisted star"
      which would lead away from the T.-T. theory).

The starting point is the field algebra ${\cal F}_N$ associated
to the ${\cal A}_N^{univ.}$-observable algebra. The latter
 is the $Z_{ZN}$ maximal extension of the Weyl-algebra
 on $S^1$ as constructed in [44].

This algebra has $2N$ central charges $C_l$ and its faithful
representation requires a Hilbert space:
\begin{eqnarray}
 {\cal H} = \mathop{\oplus}_{n=0}^{2N-1} {\cal H}_n & , & ~~~~
           C{\cal H}_n = e^{\frac{i\pi n}{N}} {\cal H}_n \\
           C = e^{\frac{2\pi iQ}{\sqrt{ 2N}}} & , & ~~~~ C_l = C^l.
\end{eqnarray}
The field algebra ${\cal F}_N$ is generated by sector-intertwining
 operators which ``live'' on a finite covering of $S^1$ and obey
 the cylinder-braid-group commutation relations of section III,
 which in this anyonic case reads as:
\begin{eqnarray}
  \phi _{\rho _1}\phi _{\rho _2} = e^{\frac{i\pi n_1n_2}{2N}}
            e^M_{mon} \phi _{\rho _2} \phi _{\rho _1}
\end{eqnarray}
with $e_{mon}=$ monodromy phase-factor, $n_i$ charge-values and
\begin{eqnarray}
&& loc \phi _{\rho_i} \subset J_i,~~~~ proj J_1 \cap proj J_2 = \oslash\\
&&J_1 + 2\pi  M < J_2 < J_1 + 2\pi  (M+1).  \nonumber
\end{eqnarray}
Here $loc$ refers to the localization region.

We restrict our discussion to the first sheet and choose
\begin{eqnarray}
  && \phi _{\rho _1} \in {\cal F}_1 = {\cal F}(S_+) ,\\
  &&  \phi _{\rho _2}
         \in {\cal F}_2 = {\cal F}(S_-).\nonumber
\end{eqnarray}
In analogy to Fermions, we  construct a twist operator $Z$
   as a suitable square root of the $2\pi $-rotation:
\begin{eqnarray}
&& e^{-2\pi iL_0} = e^{-i\pi Q^2} \nonumber \\
&& Z:= e^{-i\pi \frac{Q^2}{2}} = \sum^{2N-1}_{n=0}
           e^{-i\pi \frac{n^2}{4N}} P_n\\
&& \mbox{with} ~~ P_n = ~~ \mbox{central projector onto}~~ {\cal H}_n.
\nonumber
\end{eqnarray}
The twist transforms $\phi _{\rho _2}$ according to
\begin{eqnarray}
   Z\phi _{\rho _2} Z^{-i} & = & \phi _{\rho _2} \sum_{n}
         e^{\frac{-i\pi (n+n_2)^2}{4N}} P_n \sum e^{\frac{i\pi n^2}{4N}}
              P_n\\
   & = & e^{\frac{-i\pi n^2}{4N}} \phi _{\rho _2} V_{n_2}\\
         \mbox{with} && V_{n_2}: =
           \sum_{n} e^{-\frac{-i\pi n_2 n}{2N}}P_n.
\end{eqnarray}
As a consequence of
\begin{equation}
V_{n_2} \phi _{\rho _1} = e^{-i\frac{n_1n_2}{2N}\pi}
    \phi _{\rho _1} V_{n_2}
    \end{equation}
 this $V_{n_2}$ commutation phase precisely compensates
  the statistics phase, and we obtain for the ${\cal F}_N$-generators
  \begin{equation}
  \phi _{\rho _1} Z\phi _{\rho _2} Z^{-1} =
          Z\phi _{\rho _2} Z^{-1} \phi _{\rho _1}.
\end{equation}
 This result does not change, if the localizations
 of the $\phi _{\rho _i},~ i = 1,2$ (each in one sheet)
 are in general relative position.

 From the generator commutation
  relation we abstract the so-called ``twisted locality'', i.e.,
  the statement:
    $$
   {\cal F}(S_+) \subset {\cal F}(S_-)^q $$
   \begin{equation}
    S\pm:~~ \mbox{half~ circles}\end{equation}
   $$\mbox{with}~~~ {\cal F}(S_-)^q: = (Z{\cal F}(S_-)Z^{-1})'.
   $$
   In addition we have the involution property
   \begin{equation}
   {\cal F}(S_-)^{qq} = {\cal F}(S_-)
   \end{equation}
   since $ Z^2 = 2\pi $ rotation.

The derivation of  the twisted modularity
 from the geometrically twisted
form of causality
 for ${\cal F}(S_-)$
is well-known procedure [3] and identifies the dilation
 as the modular group.

The KMS property of the dilation manifests itself in terms
of point-like covariant fields (and only in terms of these)
as a quasiperiodicity in the analytic continuation
of their correlation functions, thus paralleling the well-known
$Dil (2\pi i)$ periodicity properties of local generators in
 ${\cal A}_N^{univ.}$.
The quasiperiodicity (as opposed to periodicity) accounts for the
braid-group commutation relations (as opposed to bosonic commutation
relations).

The modular reflection $J$ contains, as for fermions, the
twist $Z$
in addition to the TCP operator.
 The well-known modular theory of the observable
algebra ${\cal A}_N (S_+)$ in the vacuum state,
and the existence of a state-preserving conditional expection
${\cal F}\rightarrow {\cal A}$ follows
then by the ``Takesaki devissage theorem'' [31].

In ${\cal F}_N$ the charge conjugation and the star--operation
coalesce. This is the main difference of this special model
as compared to the general exchange algebra with noninteger
statistical dimensions, where a geometric Tomita-Takesaki
modular theory which is consistent
with the geometric $TCP$ theorem is still lacking.

It is however my conviction, that all models
with integer statistical dimension or (using a concept recently
developed by Rehren [45]) with vanishing
``defect projector'' can be treated with the above method.

There are arguments [46] that Hopf algebras exhaust the integer
statistical dimension sector category of properly infinite von
Neumann algebras. It has been known for some
time that Hopf $C^*$-algebras in the form of Drinfeld
``doubles'' [47] also fulfil braid-group requirements
(perhaps even the complete list of the braid-group-category axioms?).
So, such doubles may have field theoretic models with a similar
twisted geometric modular theory as above.

In this connection it is interesting to point out, that
a
framework for a
systematic study of $C^*$-Hopf-algebra categories with the additional
braid-statistics category requirements has been formulated recently [48].
\section{Selfduality and QFT-Inclusions}
For our purpose
the  most remarkable property of the matrix $S$ (i.e., the numerical values
of the various invariant charges $C_{\rho _i}$
in the irreducible representations $\pi _0 \cdot \rho _j(\cdot)$)
for our purpose
is its symmetry.

 In the case of finite nonabelian group,
the invariant charges are obtained by averaging the group representers
in the physical Hilbert-space over the various
 conjugacy classes, i.e., the index on $C_\rho $ refers to a conjugacy
class. Whereas the number of irreducible representations equals
the number of conjugacy classes, thus leading to a square matrix $S$
which is the same as the character table, the two indices remain
conceptually different (e.g., the fusion of irreducible sectors
is not the same as the fusion of conjugacy classes).

So, already in a very early state of development of low
dimensional QFT it became clear, that the new symmetry must
be more symmetric than e.g.\ finite non-abelian groups. It
 should be analogous to abelian symmetry groups
even in cases where the statistical dimensions $d_\rho $ are
bigger than one.

 There are some interesting
 lessons which one can learn from the abelian $Z_{2N}$
 model of section 3, if one analyses its duality structure from the
 modular point of view as proposed by Fredenhagen, Longo,
 Rehren and Roberts [45]. From a field algebra ${\cal F}({\cal O})$ with a
 known geometric  T.T.-modular theory and a smaller observable algebra
 ${\cal A}({\cal O})$ which is assumed to be irreducibly contained
 in ${\cal F}({\cal O})$ with finite Jones index, one may construct
 the beginning of a Jones tower (tunnel):
 \begin{equation}
 {\cal F}({\cal O})  \mathop{\supset}^{\mu } {\cal A}({\cal O})
  \mathop{\supset}^{\nu} \gamma ({\cal F}({\cal O }()
   \supset \gamma ({\cal A}({\cal O}))
     \dots
\end{equation}
 Here $\mu $ and $\nu$ are (unique) conditional expectations
 which exist according to the finite index assumption, and $\gamma $
 is Longo's canonical endomorphism which is only unique as a
 sector [49] and whose construction requires modular theory  (it may
 be represented as the product of two Tomita
 reflections belonging to ${\cal F}({\cal O})$ and
 ${\cal A}({\cal O})$). The inclusion:
 \begin{equation}
 {\cal A}({\cal O}) \supset \gamma ({\cal A}({\cal O}))
  \equiv \rho ({\cal A}({\cal O}))
\end{equation}
  is that inclusion, on which the superselection theory
  is based (to see this, one
  has to decompose $\rho $ into its
  irreducible pieces). It leads to the superselection-``paragroup''
\footnote{Ocneanu would also introduce endomorphisms for the single step
inclusion i.e. $\gamma=\sigma\bar\sigma$, $\bar\sigma$ conjugate to $\sigma$.
For a single algebra (but not coherently for the whole net) this is possible
whenever ${\cal F}(0)\sim{\cal A}(0)$. But an extension of $\sigma$ to the net
is
generally not possible.}
  (using the language of Ocneanu who first analysed
  such inclusions in the case of hyperfinite $II_1$
  algebras from the point of view of an analogue of group theory),
  whereas the paragroup going with the inclusion
   \begin{equation}
   {\cal F}({\cal O}) \supset \gamma ({\cal F}({\cal O})
  \end{equation}
   deserves the name ``symmetry'' paragroup [45]. Selfduality means that the
two paragroups behind (40) and (41) are isomorphic.

   Guido and Longo have shown that $\gamma $ allows
    to transfer the ``Mackey reduction-induction machine''
    to the sector theory of von Neumann algebras (by using
    Connes bimodules) thus generalizing earlier observation of M. Rieffel

Instead
   of giving a more detailed explanation of the various
   concepts used in these inclusions, let us specialize to the abelian
   $Z_n$-model. In that case ${\cal F}({\cal O})$ and ${\cal A}({\cal O})$
   are
   subalgebras of well-known global algebras ${\cal F}_{univ.}$
    and ${\cal A}_{univ.}$.
    In addition to localized anyonic or bosonic operators, they
    also contain global operators  $\Gamma $ (in case of
    ${\cal F}_{univ.}$) and $Q$ (in the case of ${\cal A}_{univ.}$) [43].

    $Q$ is the ``global $Z_{2N}$ charge-measurer'' and $\Gamma $
    the global ``charge-creator'':
 \begin{eqnarray}
 Q\Gamma Q^*  & = & e^{\frac{2\pi i}{N}} \Gamma\\
 \left[ Q, e^{iyL_0}\right] = 0 & , & e^{iL_0y}
                   \Gamma e^{-iyL_o} = e^{\frac{2\pi i}{N}}\Gamma .
                   \nonumber
 \end{eqnarray}
  The last relation is the transformation property of
  $\Gamma $ under rigid rotations.

The conditional expectations
  on the von Neumann algebras $\mu $ and $\nu$ may be directly
  obtained from those of the global $C^*$ algebras:
\begin{eqnarray}
   \mu (b) & = & \frac{1}{|G|} \sum_{n}Q^n bQ^{-n}~~~~~b \in {\cal F}_{univ.}\\
   \nu (a) & = & \frac{1}{|G|} \sum\Gamma ^n a\Gamma ^{-n}
              ~~~~~a \in {\cal A}_{univ.}.\nonumber
\end{eqnarray}
{}From this one obtains two global fixed point algebras
\begin{equation}
{\cal F}_{univ.} \stackrel{\mu }{\rightarrow  }
    {\cal A}_{univ.},~~~ {\cal A}_{univ.} \stackrel{\nu }{\rightarrow }
         ''\!\!\gamma ({\cal F}_{univ.})''
 \end{equation}
 The image under $\nu $ can be easily described as the extended
  Weyl algebra with the $Q$ charge removed (it is still bigger
  than the ordinary Weyl algebra on the circle!). However,
  its interpretation in terms of $\gamma $ fails (the reason for the
parenthesis),
  since $\gamma $
  can only be constructed via the T.T.-modular theory
  in case of properly infinite (e.g. III) algebras
 whereas the global algebra are type I.

  To be more precise: whereas it is possible to compute the Jones
  projector of the global (type I) inclusion ${\cal A}_{univ.} \supset
  \nu ({\cal A}_{univ.})$
  as an operator in ${\cal F}_{univ.}$ as
  \begin{equation}
   E_0 = \sum_{n} \frac{1}{ |G|} \Gamma ^n,~~~
     E_i = Q^i E_0 Q^{-i}\quad,\quad |G|=2N
\end{equation}
  with
  \begin{equation}
     \mu (E_0) = \frac{1}{|G|} 1,\end{equation}
     one cannot split $E_0$ into an isometry
     \begin{equation}
     E_0 \neq V V^*
   \end{equation}
  which is the ``would be'' generator of the Popa-Pimsner basis
  in QFT [50]
  \begin{equation}
  \mu (bV^*) V = V^* \mu (Vb) = \lambda ^{-1}b
\end{equation}
with $ \lambda = |G|$ in our case.

In order to obtain a projector $E_0$ as part of a local
(type $III_1$) algebra ${\cal F}({\cal O})$ with the
 split into Popa-Pimsner isometries $V$, one must
 replace the $\Gamma $'s by anyonic fields localized in say
 (without loss of generality) a half-circle $S_\pm$.

 But then the isometry $V$ (whose existence is secured
 by the properties of type $III$ von Neumann algebras) does not
 act in a completely local way (i.e., does not transform
 all subalgebras of ${\cal A} (A_\pm)$ into themselves as $\mu $ does).

Once the $V$
 is known, one can define a corresponding (reducible, with
 index $|G|$) canonical endomorphism $\gamma $ and a
 condition expectation $\nu $ [50]:
\begin{eqnarray}
 \gamma (b): &  = & \lambda \mu (VbV^*)~~~~b \in {\cal F} \\
 \nu (a): & = & \lambda \mu (E_0aE_0),~~~~
  \lambda =|G|~~~,~~~
  E_0 = VV^*.
\end{eqnarray}
It is easy to see that the invariance requirement $\nu (a) = a$
again imposes a linear condition on the generating charge distribution
of the extended Weyl algebras  ${\cal A}(S_\pm)$, but this
condition is less homogeneous as compared to the global
 algebra, i.e., cannot be simply expressed
as the absence of certain zero modes $Q$.

So, the self-duality picture, although by the choice of the model
true for the group theory of the system
charge-measures $Q$ and charge-creators  $\Gamma $,
  apparantly cannot be pushed
 to the local level: $\mu $ remains completely local whereas $\nu $
 is only partially local.

In contradistinction to the known ergodic behaviour of the canonical
endomorphisms constructed from double reflections of the geometric
inclusions of the second section, the inclusions of this section lead to a
Jones
tunnel with a nontrivial limiting $III_1$ algebra which in the case of the
present ${\bf Z}_{2N}$ model turns out to be a subalgebra of the circular
Weyl algebra. This remarkable distinction between ``deep inclusions" (leading
to external symmetries) and ``shallow inclusions" (yielding inner symmetries)
 clearly asks for a more profound understanding.

 It is a curious fact that the euclidean
 theory, briefly sketched in section II, allows for linearly
 rising charge distributions which are vanishing outside
 a half circle  (similar to ``blips'' [51]):
such discontinuous charge distributions
 become finite operators
with the euclidean star-operation.

 We take this as an indication
 that the euclidean quantum field theory (i.e.,
 the operator algebra formulation of the euclidean analytic continuation
 of correlation functions) may play an important role in the quest
 for a perfect (local) self-duality.

We also think of the analytic modular relation for temperature correlation
functions found in conformal models (generalizing
the modular relation between characters from which Verlinde [38]
found the relation between $S$ and the fusion matrices) as an analytic
consequence of an
algebraic local self duality obtained by using class-invariant
 temperature states. The validity of this conjecture would
 then entail that those mysterious modular identities
 from Jacobi to Ramanujan, which presently are (in most cases)
 derived by Poisson-resummation techniques, have
 their ultimate conceptual understanding via T.T.~modular theory
 in terms of the two physical principles mentioned
 in the introduction: Einstein causality and Dirac stability.\footnote{Then the
findings of Verlinde and others would return from their present
algebra-geometric
setting (not understandable by physicists) to their quantum-physical roots
(understandable by some mathematicians).}

The reader finds further speculative remarks resulting
from our formula (34) in a recent paper by Rehren. In
particular Rehren shows that nonconfined nonabelian
group-symmetries (never observed as exact symmetries in nature!)
would destroy the pretty picture of selfduality.

 \section{Speculative Remarks on ``Quantum Symmetry''}
 The observations about ${\cal G}^{inv.}$ in section~III
 and the subsequent discussions of selfduality
 invite to speculate about a full symmetry algebra ${\cal G}$.

This problem has been previously discussed in the work
of Mack and Schomerus [52] as well as Szlach\`anyi [53]
and Vecserny\`es [54]. For the conformed Ising model,
 the nontrivial $d = \frac{1}{16}$ sector appears in their global
approach with (assumed)
 multiplicity 2 in their field algebra, which then has
 an underlying weak quasi-Hopf algebra structure.
 Apart from the actual physical problem of whether such quasi-Hopf algebras,
 which lack strict associativity, can be used in practical
 calculation (symmetry breaking, Hartree-Fock approximations etc.),
 these proposals deviate in other aspects significantly from
 our picture obtained on the basis of algebraic QFT
 emphasizing ``localization'' in every step.
Actually a detailed classification of global ``rational" symmetry operations
and an understanding of their possible local shortcomings may be quite helpful.

To get an idea about ${\cal G}$, we contemplate that the overlap of the
spatial abelian $2\pi $-rotations  (only nontrivial as a covering
transformation) with the invariant charges [7] is only the
invariant projection of a bigger non-commutation overlap
 between space-time and internal symmetries.
The completely unexpected appearance of mapping
class matrices of all geni in the ${\cal A}_{univ.} (S^1)$
observable algebra lends credibility to such a conjecture.

In that case there should exist a non-commutative  subgroup
of the M\"obiusgroup (and not just of the covering) which contains
the TCP reflection and whose relative size of conjugacy classes
for the remaining generators should be measured by the Jones
numbers (thinking about minimal models).

Surprisingly, Vaughn Jones found
such a purely group theoretical quantization in which
those numbers appear based on the
following observation:
\begin{itemize}
\item[(i)]~~~
any subgroup generated by the reflection $\tilde{J} =
 \left( %
\begin{array}{rr}
    0 & 1    \\{}
    -1 & 0
\end{array}
\right) $
and a parabolic subgroup $\left( %
\begin{array}{rr}
     1 & \lambda    \\{}
     0 & 1
\end{array} \right) $
leads to quantization $\lambda = 2 \cos \frac{\pi }{q}~~~q \geq 3$
for $\lambda <2$ if the group is supposed to act in a discrete
fashion
in the sense of hyperbolic geometry
 on $R \cup \{{\infty} \}$
\item[(ii)]~~~
such groups $\Gamma $ are free group on two cyclic
 generators $\tilde{J}$ and\\  $$\left( %
\begin{array}{rr}
    1 & \lambda     \\{}
    0 & 1
\end{array}\right)  \cdot \tilde{J} =
 \left( %
\begin{array}{rr}
    -\lambda ,& 1   \\{}
    -1 & 0
\end{array}\right) =K,~~~ \mbox{with}~~K^q = 1$$
\\
with $K$ a rotation in
 $PSL(2,R)$  and $\Gamma $  the  non-amenable
 free product  $\Gamma = \tilde{J} * K$.
\end{itemize}
The group (von Neumann)-algebras of such $\Gamma $'s
are nonhyperfinite factors of type II$_1$ [55].
This means that the standard representation theory via
block-decomposition of the regular representation does not work.
So, if $\Gamma $ appears as a part of the unknown ${\cal G}$,
it should come together with some yet unknown representation
theory.

{}From the viewpoint of physical principles there is nothing
against nonhyperfinite field algebras. In
fact, the proof of hyperfiniteness only applies
to local observable algebras ${\cal A}({\cal O})$ [56].

The fact  that the field algebras of the DR
construction are also hyperfinite III$_1$ algebras is just
a fringe benefit of the permutation group statistics
which in turn results from the assumption of $4-d$
 finite energy particle states with mass gaps [58]
and has no other physical principle behind it.
Statistical
mechanics also lead to infinite hyperfinite
algebras in the thermodynamic limit (this is true by
construction).

So, the construction of physically viable
nonhyperfinite field algebras does not seem to be an easy matter.

In this context, one should mention an idea
of Voiculescu [57]. In his approach to noncommutative
probability theory based on the concept of ``freeness'',
the Bose field (at a fixed time) belongs to the standard
commutative measure theoretical situation (type I), whereas
the CAR fermionic algebra, which admits a tracial state (not
describable in measure theoretic terms), is a  hyperfinite type
II$_1$ factor. Beyond this he expects an ever-increasing wealth of
more and more non-hyperfinite algebras with more complicated
Fock-spaces (the plektons of algebraic QFT?).

In the spirit of Voiculescu one expects that the localizability
and statistical independence for space-like separation in the exchange-algebra
description ought to be replaced by space-like ``freeness" for the unknown
field algebras, similarly to the transition from statistical independence to
``freeness" in his noncommutative probability theory. I believe that the
startling
selfduality and finiteness of the new symmetry (outside abelian groups) can
only
be properly understood with the help of very big (nonhyperfinite) von Neumann
algebras: the simpler the properties, the bigger the objects carrying them.
For the physical use of such ideas one must
 of course insure that the finite region observable algebras (expected to be
the
 gauge invariant subalgebras of a generalized unknown gauge principle) remain
hyperfinite.

Last not least, the wealth of semiclassical observations made
about the new invariants of 3-manifold on the basis
of Witten's topological QFT [10], Atiyah's axiomatic
approach to this problem [59] and the Reshtkhin-Turaev-Viro combinatorial
approach [60] as well as some speculative ideas of Ocneanu [61], all
point into the same direction: an inexorable new link between space-time
and inner symmetries outwitting all the old NoGo theorems [8]
based on analytic and algebraic properties of $4-d$ QFT's!

\vskip 1cm
Part of the work has been carried out at the Berkeley Mathematics Department
as well as during a short visit of the DESY Theory Group. I gratefully
acknowledge the hospitality of Vaughn Jones and Martin L\"uscher.
\newpage

\noindent {\Large\bf References}
\medskip

\begin{enumerate}
\item R. Haag, Local Quantum Physics, Springer 1992.
\item D. Buchholz and P. Junglas "Commun.Math.Phys.{\bf 121}, 255 (1989).
\item Articles of J.E. Roberts, D. Kastler, and K.H. Rehren in:
The Algebraic Theory of Superselection Sectors. D. Kastler (ed.),
Singapore, World Scientific 1990,
as well as ref. [1].
\item J.E. Roberts in $C^*$-algebras and their applications to statistical
mechanics
and quantum field theory, ed. D. Kastler, North Holland (1976).\\
K. Fredenhagen, to appear in the Proceedings of the Carg\`ese Summer Institute
(1991).
\item K. Fredenhagen, K.H. Rehren, and B. Schroer, Commun.Math.Phys.{\bf
125}\\
(1983),201.
\item S. Doplicher, R. Haag,  and J.E. Roberts, Commun.Math.Phys.{\bf
23}(1971),149 and
{\bf 35}(1974)49.
\item K. Fredenhagen, K.H. Rehren, and B. Schroer, Reviews in Mathematical
Physics,
Special Issue (1992) 113.
\item S. Coleman and J. Mandula, Phys.Rev.{\bf 159}, 1251(1967).\\
L. O'Raifeartaigh, Phys.Rev.Lett{\bf14},575(1965).\\
W.D. Garber and H. Reeh, J.Math.Phys.{\bf 19},985(1978).
\item Remarks on the construction 3-manifolds in the frameworks of algebraic
QFT,
their relations with the Reshetikhin-Turaev resp. Turaev-Viro combinatorical
approach, as well as
and their possible physical interpretation in terms of properties of plektons
can
be found in the appendix of reference [7].
\item E. Witten, Commun.Math.Phys.{\bf 121}(1989)351.
\item F.Nill, International Journal of Mod.Phys. B{\bf 6}(1992)2159.\\
In order to obtain the topological field theory of the $Z_{2N}$ model i.e. the
centralizer algebra (14), one has to extend the Weyl-algebra of 1-forms by
the universal construction which leads to closed but not necessarily exact
forms
(in the angular variable). The latter construction is therefore a kind of
algebraic
compactification.
\item H. Narnhofer and W. Thirring, Rev.in Math.Phys., Special Issue (1992).
Here the general idea of using singular states for obtaining gauge invariant
subalgebras
is developed.
\item For the Bargman Hall-Wightman theorem on the analytic extension of vacuum
expectatin values of point-like covariant fields we refer to \\
R.F. Streater and A.W. Wightman: PCT, Spin and Statistics and all That,
Benjamin, New York (1964).
\item B. Schroer, Nucl.Phys. B{\bf 369} (1992), 478.
\item M. Karowski, H.-J. Thun, T.T. Truong and P.H. Weisz, Phys.Lett.{\bf
67B},321 (1977).
\item A.B. Zamolodchikov, JETP Lett.{\bf 499} (1977).
\item D. Buchholz and K. Fredenhagen, J.Math.Phys.{\bf 18},5 (1977).
\item M. Karowski and P. Weisz, Nucl.Phys. {\bf B139} (1979) 209107.\\
F.A. Smirnov, Nucl.Phys.{\bf B337}(1990)\\
For a recent account with an interesting physical application see also
J.Cardy and G. Mussardo ``Universal Properties of Self-Avoiding Walks from
Two-Dimensional Field Theory" ISAS preprint 1993.
\item B. Schroer and J.A. Swieca, Nucl.Phys.{\bf B121} (1977) \\
For a formulation in the setting of algebraic QFT see K. Fredenhagen [34] and
K. Fredenhagen, work in progress.
\item K. Fredenhagen, M. Gaberdiel and S.M. R\"uger, ``Scattering States of
Plektons in (2+1) Dimensional Quantum Field Theory" in preparation.
\item See Theorem 4.15 of [13].
\item V. Schomerus, Quantum symmetry in quantum theory, DESY preprint 1993\\
and references therein.
\item V. Kac and M. Wakimoto, Adv.Math.{\bf 70} (1986).
\item H.J. Borchers, Commun.Math.Phys.{\bf 143} (1992) 315.
\item B. Schroer, Int.J. of Mod.Phys. {\bf B}, 6 (1992) 2041.\\
Formula (11) and (12) in that paper contain an unfortunate misprint.
${\cal A}(I)$ should be replaced by ${\cal L}(I)$.
\item H.W. Wiesbrock, Lett. in Mod.Phys. {\bf 28},107 (1993).\\
Halfsided modular inclusions of von Neumann algebra, Commun.Math.Phys., in
press.
\item D. Buchholz and H. Schulz-Mirbach, Reviews in Math.Phys.{\bf 2}, 1 (1990)
105.
\item J. Fr\"ohlich and F. Gabbiani, Operator algebra and conformal field
theory
ETH Z\"urich preprint (1992).
\item R. Brunnetti, D. Guido and R. Longo, Modular structure and duality in
conformal
field theory, University of Rome II, preprint (1992).
\item The ``split property" goes back to Borchers and Doplicher-Longo, see [1].
\item A. Wassermann, ``Subfactors arising from positiv-energy representation of
some infinite dimensional groups", unpublished notes 1990.
\item See the contribution of H. Araki and A. Connes to ``$C^*$-Algebra and
their
Applications to Statistical Mechanics and Quantum Field Theory", Proceedings of
the
International School of Physics ``Enrico Fermi", Course IX, Varenna 1973,
Intalian Physical Society.
\item Joint work with H.W. Wiesbrock, in preparation. In incomplete account of
this
work can be found in [25].
\item K. Fredenhagen in ``The Algebraic Theory of Superselection Sectors",
D. Kastler (ed.), Singapore, World Scientific 1980.
\item K.H. Rehren and B. Schroer, Phys.Lett {\bf 198B}, 84 (1987).\\
Nucl.Phys. {\bf B321}, 3 (1989).
\item K.H. Rehren, Commun.Math.Phys.{\bf 145} (1992) 123.
\item V.S. Sunder, J. Operator Theory {\bf 18} (1987) 289.\\
These graphical methods for inclusion theory were first used in\\
A. Ocneanu, unpublished, Warwick lecture notes (1987).\\
Within the DHR theory they were  used in [5].
\item E. Verlinde, Nucl.Phys.{\bf 300}, 360 (1988).
\item G. Moore and N. Seiberg, Phys.Lett.{\bf B212}, 451 (1988).
\item K.H. Rehren in: ``The Algebraic Theory of Superselection Sectors,
D. Kastler (ed.), Singapore, World Scientific 1990.
\item M. Karowski and R. Schrader, Commun.Math.Phys.{\bf 151}, 355 (1993),
especially Fig. 25.
\item S. Doplicher and J.E. Roberts, Commun.Math.Phys.{\bf 131} (1990) 51.
\item K. Dr\"uhl, R. Haag and J.E. Roberts, Commun.Math.Phys.{\bf 18} (1970)
204.\\
The step from parafields to the exchange algebra (reduced field bundle) with
R-matrix commutation relation $(R ^2=1)$ is scetched in my contribution to
The Algebraic Theory of Superselection Sectors, D. Kastler (ed.), Singapore,
World Scientific 1990.
\item D. Buchholz, G. Mack, and I. Todorov, Nucl.Phys. B (Proc. Suppl.) {\bf
53}
(1988) 20.
\item K.H. Rehren, Subfactors and Coset Models, DESY preprint August 1993.\\
K. Fredenhagen, R. Longo, K.H. Rehren, J.E. Roberts, work in progress.
\item R. Longo, a  duality for Hopf algebras and for subfactors I, Univ. of
Rome II,
preprint 1992.
\item S. Baaj and G. Skandalis, Unitaires multiplicatifs et dualit\'e pour les
produits crois\'es de $C^*$ alg\`ebres, preprint 1991.\\
J. Cuntz, Regular actions of Hopf algebras on the $C^*$-algebra generated by a
Hilbert
space, preprint 1991.
\item T. Ceccherini, S. Doplicher, C. Pinzari and J.E. Roberts, A gneralization
of the Cuntz algebras and model action, University of Rome I, preprint 1993.
\item D. Guido and R. Longo, Commun.Math.Phys.{\bf 150}, 521 (1992),\\
and Longo's previous work on the canonical endomorphism quoted therein.
\item The field-theoretic formulation as well as the reference to the general
construction by Popa and Pimsner is contained in [45].
\item A. Pressley and G. Segal, Loop Groups, Oxford University Press (1986).
\item G. Mack and V. Schomerus, Commun.Math.Phys.{\bf 137} (1990) 57.\\
V. Schomerus, Quantum symmetry in quantum theory, DESY preprint 1993.
\item K. Szlach\`anyi, Chiral decompositon as a source of quantum symmetry in
the
Ising model, University of Budapest preprint 1993.
\item P. Veccerny\`es, On the quantum symmetry of the chiral Ising Model,
Princeton Unversity, preprint 1993.
\item An account of this Jones quantization which embeds the Hecke-groups
inside
the M\"obiusgroup can be found in F.M. Goodman, P. de la Harpe and V.F.R.
Jones,
Coxeter Graphs and Towers of Algebras, Springer 1989. Appendix III.
\item D. Buchholz, C. D'Antoni and K. Fredenhagen, Commun.Math.Phys.{\bf 111}
 (1987) 123.
\item See the introduction of a new book by D. Voiculesu, Free Probability
Theory,
Bures Sur Yvette 1992.
\item D. Buchholz and K. Fredenhagen, Commun.Math.Phys.{\bf 84}, 1 (1982).
\item M. Atiyah, Publ.Math.Inst. Hautes Etudes Sci. Paris {\bf 68}, 175-186
(1989).
\item N. Yu. Reshetikhin and V.G. Turaev, Inv.Math.{\bf 103}, 547 (1991).\\
V.G. Turaev and O.Y. Viro, to appear in Topology.
\item A. Ocneanu, unpublished, private communication.
\item M. Karowski, R. Schrader and E. Vogt, ``Unitary representations of the
mapping class group and numerical calculations of invariants of hyperbolic
three
manifolds", in preparation.
\item D. Buchholz and K. Fredenhagen, Commun.Math.Phys.{\bf 84}, 1 (1982).
\item B. Schroer, Phys.Lett.B{\bf 199}, 183 (1987).
\end{enumerate}

\end{document}